\documentclass[12pt,a4paper]{article}
\pdfoutput=1
\usepackage{graphicx}
\usepackage[T1]{fontenc}
\usepackage[sc,osf]{mathpazo}
\usepackage{a4wide}  
\usepackage{latexsym,amsthm,amsfonts,amsmath,mathrsfs,amssymb}
\usepackage[unicode,implicit]{hyperref}
\hypersetup{%
  pdftitle    = {Yang-Mills instantons in Kahler spaces with one holomorphic isometry}
  pdfkeywords = {Yang-Mills, instanton,  monopole, Kahler, holomorphic},
  pdfauthor   = {Samuele Chimento, Tomas Ortin, Alejandro Ruiperez},
  plainpages  = true,
  colorlinks  = true,
  citecolor   = blue,
  urlcolor    = red,
  linkcolor   = black
}
\newcommand{\hepth}[1]{{\tt
\href{http://www.arXiv.org/abs/hep-th/#1}{hep-th/#1}}}

\newcommand{\arxiv}[1]{{\tt arXiv:\href{http://www.arXiv.org/abs/#1}{#1}}}
\newcommand{\doi}[1]{{\tt DOI:\href{http://dx.doi.org/#1}{#1}}}

\makeatletter
\@addtoreset{equation}{section}
\makeatother

\begin{document}

\begin{flushright}
\small

IFT-UAM/CSIC-17-076\\
October 2\textsuperscript{nd}, 2017\\
\normalsize
\end{flushright}

\vspace{1cm}

\begin{center}
 
{\Large {\bf Yang-Mills instantons in K\"ahler spaces\\[.5cm] with one holomorphic isometry}}
 
\vspace{1.5cm}

\renewcommand{\thefootnote}{\alph{footnote}}
{\sl Samuele Chimento}\footnote{E-mail: {\tt Samuele.Chimento [at] csic.es}},
{\sl Tom\'{a}s Ort\'{\i}n}\footnote{E-mail: {\tt Tomas.Ortin [at] csic.es}}
{\sl and Alejandro Ruip\'erez}\footnote{E-mail: {\tt alejandro.ruiperez [at] uam.es}},

\setcounter{footnote}{0}
\renewcommand{\thefootnote}{\arabic{footnote}}

\vspace{1.5cm}

{\it Instituto de F\'{\i}sica Te\'orica UAM/CSIC\\
C/ Nicol\'as Cabrera, 13--15,  C.U.~Cantoblanco, E-28049 Madrid, Spain}\\

\vspace{1.5cm}


{\bf Abstract}

\end{center}

\begin{quotation}
  We consider self-dual Yang-Mills instantons in 4-dimensional K\"ahler spaces
  with one holomorphic isometry and show that they satisfy a generalization of
  the Bogomol'nyi equation for magnetic monopoles on certain 3-dimensional
  metrics. We then search for solutions of this equation in 3-dimensional
  metrics foliated by 2-dimensional spheres, hyperboloids or planes in the
  case in which the gauge group coincides with the isometry group of the
  metric (SO$(3)$, SO$(1,2)$ and ISO$(2)$, respectively). Using a generalized
  hedgehog ansatz the Bogomol'nyi equations reduce to a simple differential
  equation in the radial variable which admits a universal solution and, in
  some cases, a particular one, from which one finally recovers instanton
  solutions in the original K\"ahler space. We work out completely a few
  explicit examples for some K\"ahler spaces of interest.
\end{quotation}

\newpage
\pagestyle{plain}



\section{Introduction}

There is a well-known relation between Yang-Mills instantons in four flat
Euclidean dimensions and static magnetic monopoles of the Yang-Mills-Higgs
theory in four flat Lorentzian dimensions. A particular case that has focused most
of the research is the relation between (anti-) selfdual instantons and
monopoles satisfying the Bogomol'nyi equation \cite{Bogomolny:1975de} in
$\mathbb{E}^{3}$.\footnote{The time coordinate is irrelevant in this problem
  because of the restriction to static monopoles.} These are usually known as
BPS magnetic monopoles because the SU$(2)$ 't~Hooft-Polyakov monopole
\cite{tHooft:1974kcl,Polyakov:1974ek} satisfies it in the Prasad-Sommerfield
limit \cite{Prasad:1975kr}. Both the (anti-) selfduality condition and the
Bogomol'nyi equation are first-order equations that imply that the action is
extremized locally and the second-order Euler-Lagrange equations are
automatically satisfied. On the one hand, first-order equations are easier to
solve than second-order ones, and, for instance, this allowed Protogenov to
construct all the spherically-symmetric SU$(2)$ BPS magnetic monopoles using
the so-called \textit{hedgehog ansatz}, that exploits the relation between the
isometry group of the solution and the gauge group
\cite{Protogenov:1977tq}. On the other, these, as many other interesting
first-order equations, naturally arise in the context of supersymmetric
theories, when one searches for field configurations preserving some unbroken
supersymmetries. Supersymmetric solutions have many interesting properties,
which makes them worth studying for their own sake.\footnote{For a review on
  field configurations and solutions with global or local unbroken
  supersymmetries see, for instance, Ref.~\cite{Ortin:2015hya}.}  The
possibility of constructing solutions, such as the one in
Ref.~\cite{Ramirez:2016tqc}, describing an arbitrary number of magnetic
monopoles in static equilibrium is one of the most remarkable ones, and has
been exploited to construct dyonic, non-Abelian multi-black-hole solutions in
4-dimensional Super-Einstein-Yang-Mills theories \cite{Meessen:2017rwm}.

In Ref.~\cite{kn:KronheimerMScThesis} Kronheimer showed that the above
relation between (anti-) selfdual instantons in $\mathbb{E}^{4}$ and monopoles
satisfying the Bogomol'nyi equation in $\mathbb{E}^{3}$ could be extended to a
relation between (anti-) selfdual instantons in 4-dimensional hyperK\"ahler
spaces admitting a triholomorphic isometry, usually known as Gibbons-Hawking
spaces \cite{Gibbons:1979zt,Gibbons:1979xm}, and, again, monopoles satisfying
the Bogomol'nyi equation in $\mathbb{E}^{3}$. This map between instantons and
monopoles is surjective and from a given BPS monopole solution one can
construct an instanton solution in every Gibbons-Hawking space, which is
characterized by an additional function $H$, harmonic in $\mathbb{E}^{3}$,
which is not part of the monopole fields.  Thus, one can use all the
spherically-symmetric SU$(2)$ BPS magnetic monopoles found by Protogenov to
construct instantons with SO$(3)$ symmetry in any Gibbons-Hawking space, for
instance.\footnote{The relation does not preserve the regularity of the
  solutions in either sense. In particular, the well-known BPST SU$(2)$
  instanton \cite{Belavin:1975fg} on $\mathbb{E}^{4}_{0}$, characterized by
  the choice $H=1/r$, gives rise to the Protogenov solution known as
  \textit{coloured monopole}, which is singular \cite{Bueno:2015wva}. On the
  other hand, the rest of the spherically-symmetric BPS monopoles (including
  the globally regular 't~Hooft-Polyakov one) give rise to badly-behaved
  instantons solutions in the same GH space and one must consider other GH
  spaces \cite{CMOR}.}

This mechanism has been used in the context of the construction of timelike
supersymmetric solutions of 5-dimensional Super-Einstein-Yang-Mills theories
(non-Abelian black holes and rings, microstate geometries and global
instantons
\cite{Meessen:2015enl,Ortin:2016bnl,Ramirez:2016tqc,Cano:2017sqy,Cano:2017qrq})
because the spacetime metrics of these 5-dimensional supersymmetric solutions
are constructed using a 4-dimensional hyperK\"ahler metric (often called
``base space metric'') in terms of which a piece of the 2-form field strengths
of the theory is forced by supersymmetry to be selfdual
\cite{Bellorin:2007yp}.

In 5-dimensional supergravities with Abelian gaugings via Fayet-Iliopoulos
terms the base-space metric is forced to be K\"ahler in supersymmetric
solutions \cite{Gauntlett:2003fk} and, if there are additional non-Abelian
gaugings of the isometries of the real Special scalar manifold
(\textit{i.e.}~we are dealing with a Super-Einstein-Yang-Mills theory with an
additional Abelian gauging that introduces a non-trivial scalar potential,
among other things), one faces the problem of finding selfdual instantons in
K\"ahler spaces.\footnote{This is not just a very twisted academic problem:
  this kind of theories arise naturally in Type~II Superstring
  compactifications to 5 dimensions \cite{ChOR}.}

Just as in the hyperK\"ahler case, it is convenient to have a
``parametrization'' of the class of metrics under consideration in order to
find a set of differential equations for the problem. The space of
hyperK\"ahler metrics is very large and finding a generic one for a
hyperK\"ahler metric, in terms of a small number of functions satisfying some
relations is too complicated or impossible. The restriction to GH metrics,
which depend on just one independent function, transforms the selfduality
condition into a set of differential equations which, in the end, can be
identified with the Bogomol'nyi equations. Alternatively, one can just view
the requirement of the existence of a triholomorphic isometry as a condition
necessary to dimensionally reduce the equations along the isometric direction
preserving the hyperK\"ahler structure.

In the K\"ahler case it is natural to assume the existence of a holomorphic
isometry along which a dimensional reduction can be performed preserving the
K\"ahler structure (and supersymmetry as well, in the supersymmetric
context). In Ref.~\cite{Chimento:2016run} and references therein, it was shown
that these metrics can be written in terms of essentially two real functions
related by a differential equation (see Eq.~(\ref{eq:final_metric})) and in
this paper we are going to make use of this result to transform the
selfduality equations of a Yang-Mills field on these metrics into a set of
differential equations which ultimately can be seen as a generalization of the
Bogomol'nyi equations in some 3-dimensional space
(Section~\ref{sec-generalizedBeqs}). 

While the physical interpretation of these 3-dimensional equations is not as
transparent as those obtained in the hyperK\"ahler space (in particular, it is
not clear that they correspond to BPS magnetic monopole solutions in general),
they provide an excellent starting point to construct instanton solutions.
Thus, in Section~\ref{sec-generalizedhansatz} we are going to use the hedgehog
ansatz and generalizations thereof adequate for other gauge groups in order to
simplify the equations and obtain explicit instanton solutions in some simple
K\"ahler spaces of interest in gauged 5-dimensional supergravity, such as
$\overline{\mathbb{CP}}^{2}$.  Section~\ref{sec-conclusions} contains our
conclusions.

\section{Generalized Bogomol'nyi equations}
\label{sec-generalizedBeqs}

Any 4-dimensional K\"ahler metric admitting a holomorphic isometry can be
written as \cite{Chimento:2016run}

\begin{equation}
\label{eq:final_metric} 
ds_{4}^{2} 
= 
H^{-1}\left( dz+\chi \right)^{2}
+H\left\{(dx^{2})^{2}+W^{2}(\vec{x})[(dx^{1})^{2}+(dx^{3})^{2}]\right\}\, ,
\end{equation} 

\noindent
with the functions $H$ and $W$, and the 1-form $\chi$, independent of $z$ and
satisfying the constraint\footnote{Underlined indices refer to the coordinate
  basis.}

\begin{equation}
\label{eq:constraintijcurved}
\breve{\star}_{3} d\chi =dH+H\partial_{\underline{2}} \log W^{2} dx^{2}\, ,
\end{equation}

\noindent
where $\breve{\star}_{3}$ is the Hodge dual in the 3-dimensional manifold 

\begin{equation}\label{eq:tridimensionalmetric}
d\breve{s}^{2}_{3}
=
(dx^{2})^{2}+W^{2}(\vec{x})[(dx^{1})^{2}+(dx^{3})^{2}]\, .
\end{equation}

\noindent
The integrability condition of this constraint is a $W$-dependent deformation
of the Laplace equation for $H$ on $\mathbb{E}^{3}$

\begin{equation}
\label{eq:integrability}
\partial_{\underline{1}}\partial_{\underline{1}} H
+\partial_{\underline{2}}\partial_{\underline{2}}(W^{2} H)
+\partial_{\underline{3}}\partial_{\underline{3}} H 
= 0\, .
\end{equation}

Thus, we can construct K\"ahler metrics with a holomorphic isometry by
choosing some function $W$, solving the above integrability condition for $H$
and then solving the constraint Eq.~(\ref{eq:constraintijcurved}) for the
1-form $\chi$.\footnote{We can also construct metrics of this kind starting
  with an arbitrary real function $\mathcal{K}(x^{1},x^{2},x^{3})$ and
  computing directly
\begin{equation}
\begin{array}{rclrcl}
H 
& = & 
\partial_{\underline{2}}^{2} \mathcal{K}\, ,
\hspace{1cm}
&
W^{2} 
& = &
-H^{-1} 
\left(\partial_{\underline{1}}^{2}+\partial_{\underline{3}}^{2}
\right)\mathcal{K}\, ,
\\
& & & & & \\
\chi_{\underline{1}} 
& = &
-\partial_{\underline{3}}\partial_{\underline{2}}
\mathcal{K}\, ,
\hspace{1cm}
&
\chi_{\underline{3}} 
& = &
\partial_{\underline{2}}\partial_{\underline{1}}
\mathcal{K}\, ,
\end{array}
\end{equation}
which solve all the above equations in coordinates in which 
$\chi_{\underline{2}}=0$. 
} 
Observe that the choice $W=1$ yields hyper-K\"ahler metrics
with a triholomorphic isometry, also known as Gibbons-Hawking metrics
\cite{Gibbons:1979zt,Gibbons:1979xm}.

We are interested in Yang-Mills fields $A^{I}$ in the above space which are
$z$-independent (at least in some gauge) and whose 2-form field strengths

\begin{equation}
F^{I}
=
d A^{I}+\tfrac{1}{2}g f_{JK}{}^{I} A^{J}\wedge  A^{K}\, ,
\end{equation}

\noindent
are self-dual  

\begin{equation}
\label{eq:selfduality}
F^{I}= +\star_{4} F^{I}\, . 
\end{equation}

\noindent
Here $\star_{4}$ is the Hodge operator in the full 4-dimensional metric
Eq.~(\ref{eq:final_metric}), with the orientation $\epsilon^{z123}=+1$.

Following Kronheimer, who considered the hyper-K\"ahler case ($W=1$) in
Ref.~\cite{kn:KronheimerMScThesis}, we decompose $A^{I}$ as

\begin{equation}\label{eq:instanton}
A^{I}=-H^{-1}\Phi^{I} (dz+\chi)+\breve{A}^{I}\, ,
\end{equation}

\noindent
and substituting into the self-duality equation (\ref{eq:selfduality}) one
finds that it is equivalent to the following generalization of the Bogomol'nyi
equation \cite{Bogomolny:1975de}\footnote{The standard Bogomol'nyi equation is
  defined in Euclidean 3-dimensional space.}

\begin{equation}
\label{nogomolny}
\breve{\star}_{3} \breve{F}^{I}-\breve{\mathfrak{D}}\Phi^{I}
=
\Phi^{I} \partial_{\underline{2}}
\log W^{2} dx^{2}\, ,
\end{equation}

\noindent
where the $\breve{~}$ sign in the field strength and the covariant derivative
refers to the 3-dimensional Yang-Mills connection $\breve{A}^{I}$.

In general, the right-hand side of this equation does not have a clear
geometric or field-theoretic meaning and, therefore, there is no obvious
relation between the equation and the Yang-Mills-Higgs action in some
4-dimensional spacetime: unlike what happens with the usual Bogomol'nyi equation, it
does not seem to be related to the extremization of an action of this kind and
it does not guarantee that the corresponding second order Yang-Mills-Higgs
equations of motion are satisfied. As explained in the introduction, this does
not make them completely useless or meaningless, because they can arise in
more complex theories such as 5- and 4-dimensional gauged supergravities.

Many of the most interesting K\"ahler metrics in this class are characterized by a
function $H$ that only depends on $x^{2}$, which we will denote by $\varrho$
from now on. Equation (\ref{eq:integrability}) implies that $W$ is of the form
\cite{Chimento:2016mmd}

\begin{equation}
W^{2}(\vec{x})=\frac{\varrho}{H}\Phi_{1}(x^{1},x^{3})
+\frac{1}{H}\Phi_{2}(x^{1},x^{3})\, .
\end{equation}

\noindent
We will consider, for the sake of simplicity, the case in which either
$\Phi_{1}=0$ or $\Phi_{2}=0$ so that, calling the surviving function $\Phi$,
$W^{2}$ can be written as

\begin{equation}
W^{2} = \Psi(\varrho)\Phi(x^{1},x^{3})\, ,
\,\,\,\,\,
\mbox{where}
\,\,\,\,\,
\Psi(\varrho) \equiv \frac{\varrho^{\epsilon}}{H(\varrho)}\, .
\end{equation}

\noindent
Thus, the metrics in this class are completely determined by an arbitrary
function of $\varrho$ (either $\Psi$ or $H$) and an arbitrary function $\Phi$ of
$x^{1},x^{3}$.

For these metrics, the Bogomol'nyi equation (\ref{nogomolny}) takes the more
geometric expression

\begin{equation}
\label{nogomolnysimplified}
\Psi\breve{\star}_{3} \check F^{I}
-\breve{\mathfrak{D}}(\Psi\Phi^{I})
=
0\, ,
\end{equation}

\noindent
which can be derived from the Yang-Mills-Higgs action with a Higgs field
$\tilde{\Phi}^{I}=\Psi\Phi^{I}$ in a $1+3$ (spacetime) metric of the form

\begin{equation}
d\tilde{s}^{2}_{1+3}=g_{tt}\, dt^{2} -\Psi^{2}(\varrho)d\breve{s}^{2}_{3}\, ,  
\end{equation}

\noindent
for any time-independent $g_{tt}$ by the usual squaring of the action
arguments. 

We are not going to follow this line of reasoning any further here. Instead,
we will just try to find some explicit solutions to the above Bogomol'nyi
equation and the corresponding instantons for some interesting K\"ahler
metrics.

\section{Generalized hedgehog ansatz and solutions}
\label{sec-generalizedhansatz}

As a further simplification, we are going to restrict ourselves to the case in
which the 2-dimensional metric $\Phi(x^{1},x^{3})[(dx^{1})^{2}+(dx^{1})^{3}]$
is maximally symmetric. The three distinct possibilities, namely the round
sphere $S^{2}$, the hyperbolic plane $\mathbb{H}_{2}$ and the Euclidean plane
$\mathbb{E}^{2}$, are encompassed by the function

\begin{equation}
\Phi_{(k)}(x^{1},x^{3})
=
\frac{4}{\left\{1+k[(x^{1})^{2}+(x^{3})^{2}]\right\}^{2}}\, , 
\end{equation}

\noindent
for the values of the parameter $k$ being $+1$, $0$ and $-1$, respectively.  For
this particular kind of metrics, the 1-form $\chi$ that occurs in the metric
Eq.~(\ref{eq:final_metric}) is given by 

\begin{equation}
\label{eq:chik-1}
\chi = \epsilon\chi_{(k)}\, ,
\,\,\,\,\,
\mbox{with}
\,\,\,\,\,
\chi_{(k)}= \frac{2(x^{3}dx^{1}-x^{1}dx^{3})}{1+k[(x^{1})^{2}+(x^{3})^{2}]}\, .  
\end{equation}

The
generic coordinate change\footnote{The $k=0$ case should be seen as the limit
  $k\rightarrow 0$. 
  }

\begin{equation}
x^{1}
=
k^{-1/2}\tan{(k^{1/2}\theta/2)}\cos{\varphi} \, ,
\quad 
x^{2}
=
\varrho\, ,
\quad
x^{3}
=
k^{-1/2}\tan{(k^{1/2}\theta/2)}\sin{\varphi}\, ,
\end{equation}

\noindent
brings the 3-dimensional metric (\ref{eq:tridimensionalmetric}) into the
form

\begin{equation}
\label{eq:dOk}
d\breve{s}^{2}_{3} = d\varrho^{2}+\Psi(\varrho)d\Omega^{2}{}_{(k)}\, ,
\,\,\,\,\,
\mbox{where}
\,\,\,\,\,
d\Omega^{2}{}_{(k)}  = d\theta^{2}+k^{-1}\sin^{2}{(k^{1/2}\theta)} d\varphi^{2}\, , 
\end{equation}

\noindent
and $\chi_{(k)}$ to the form

\begin{equation}
\label{eq:chik-2}
\chi_{(k)} = k^{-1}[\cos{(k^{1/2}\theta)}-1]d\varphi\, .  
\end{equation}

It is natural to search for monopole solutions in gauge groups which coincide
with the isometry group of the maximally symmetric 2-dimensional spaces that foliate the
3-dimensional metrics that we are considering here: SO$(3)$, ISO$(2)$ and
SO$(1,2)$, respectively. Observe that, while the non-semisimple group ISO$(2)$
is just a mere curiosity, the non-compact group SO$(1,2)$ actually occurs in
supergravity theories without any of the pathologies that arise in
Yang-Mills(-Higgs) theories because these theories have scalar-dependent
kinetic matrices which make compatible SO$(1,2)$ symmetry with
positive-defined kinetic energies.

For instance, in $\mathcal{N}=1,d=5$ supergravities, the vector fields kinetic
terms are of the form

\begin{equation}
\sim a_{IJ}(\phi)F^{I}\wedge \star F^{J}\, .  
\end{equation}

\noindent
In theories with SO$(1,2)$ symmetry the scalar fields and, hence, the kinetic
matrix $a_{IJ}(\phi)$ transform under that group so that this kinetic term is
positive definite and invariant. For the kind of field configurations that we
are considering (selfdual instantons living in some 4-dimensional Euclidean
submanifold) the contribution to the action of this term would be a
positive-definite generalization of the usual instanton number $\sim
a_{IJ}(\phi)F^{I}\wedge F^{J}$.

In these theories, there is another term relevant for this discussion: the
r.h.s.~of the equations of motion of the vector fields contains a term of the
form

\begin{equation}
\sim C_{IJK}F^{I}\wedge F^{J}\, ,  
\end{equation}

\noindent
where $C_{IJK}$ is a constant, symmetric tensor, invariant under the gauge
group. This term will contain the SO$(1,2)$ Killing metric and will be
proportional to the non-definite positive ``instanton number'', but the sign
of these terms is irrelevant for the consistency of the theory and the
SO$(1,2)$ selfdual instanton solutions  are of potential interest in
consistent theories.

For the gauge group SO$(3)$, the so-called \textit{hedgehog ansatz} leads to
the construction of all the magnetic monopole solutions with this geometry
\cite{Protogenov:1977tq}. Here we propose a generalization of this ansatz that
encompasses the three cases we are considering. In terms of the structure
constants $f_{IJ}{}^{K}$ of these groups\footnote{For $k=\pm 1,0$ the
  structure constants are given by $f_{IJ}{}^{K}= \varepsilon_{IJL}\eta^{LK}$,
  where $(\eta^{IJ})\equiv \left(\begin{smallmatrix}
      1&0&0\\
      0&k&0\\
      0&0&1\\
\end{smallmatrix}\right)$.} and the coordinates $y^{I}=y^{I}(\varrho,
\theta,\varphi)$ 

\begin{equation}
\left\{
\begin{aligned}
y^{1}
& = 
\varrho\, k^{-1/2}\sin{(k^{1/2}\theta)}\cos{\varphi}\, , 
\\
y^{2} 
& =
\varrho \cos{(k^{1/2}\theta)}\, , 
\\
y^{3} 
& = 
\varrho\, k^{-1/2}\sin{(k^{1/2}\theta)}\sin{\varphi}\, ,
\end{aligned}
\right.
\hspace{1cm}
\Rightarrow
\hspace{1cm}
(y^{2})^{2} +k[(y^{1})^{2}+(y^{3})^{2}]=\varrho^{2}\, ,
\end{equation}

\noindent
the generalized hedgehog ansatz for the Higgs and Yang-Mills fields can be
written in the form

\begin{equation}
\label{eq:hedgehogansatz}
\begin{aligned}
\Phi^{I} 
& = 
F(\varrho) \frac{y^{I}}{\varrho}\, ,
\\
\breve{A}^{I} 
& = 
J(\varrho) f_{JK}{}^{I}\frac{y^{J}}{\varrho}d\left(\frac{y^{K}}{\varrho}\right)\, ,
\end{aligned}
\end{equation}

\noindent
where $F(\varrho)$ and $J(\varrho)$ are two functions to be determined. 

Plugging the ansatz into the Bogomol'nyi equation (\ref{nogomolnysimplified}),
we get the following system of first-order differential equations involving
$F,J$ and the metric function $\Psi$:

\begin{equation}
\label{eq:system1}
\begin{aligned}
\left(\Psi F\right)' 
& = 
kJ\left(2+gJ\right)\, , 
\\
J'
& = 
F\left(1+gJ\right)\, . 
\end{aligned}
\end{equation}

\noindent
where primes stand for derivatives with respect to $\varrho$.

In Section~\ref{sec-solutions} we are going to search for explicit solutions of
this system, but, before, we are going to show the form of the 4-dimensional
instanton fields in terms of the functions that appear in these equations and
in the K\"ahler metric.

\subsection{Instanton fields}
\label{sec-instantonfields}

Given a solution of Eqs.~(\ref{eq:system1}), $F(\varrho),J(\varrho)$ for a
K\"ahler metric characterized by the function $\Psi(\varrho)$ and the
parameters $\epsilon=0,1$ and $k=\pm 1,0$

\begin{equation}
\label{eq:particularmetric} 
ds_{4}^{2} 
= 
\frac{\Psi}{\varrho^{\epsilon}}(dz+\epsilon\,\chi_{(k)})^{2}
+\frac{\varrho^{\epsilon}}{\Psi}d\varrho^{2}
+\varrho^{\epsilon}d\Omega^{2}_{(k)}\, ,
\end{equation} 

\noindent
with $d\Omega^{2}_{(k)}$ and $\chi_{(k)}$ given by Eq.~(\ref{eq:dOk}) 
and Eq.~(\ref{eq:chik-2}), respectively, the instanton field is given by 

\begin{equation}
\begin{aligned}
A^{1}
= &
-\frac{\Psi F}{\varrho^{\epsilon}}k^{-1/2}\sin{(k^{1/2}\theta)}\cos{\varphi} 
[dz+\epsilon \chi_{(k)}]
\\ 
&
\\
&
+J
\left[
\sin{\varphi} \: d\theta
+ k^{-1/2}\sin{(k^{1/2}\theta)}\cos(k^{1/2}\theta)\cos{\varphi}\: d\varphi
\right]\, ,
\\
& 
\\
A^{2}
= &
-\frac{\Psi F}{\varrho^{\epsilon}}\cos (k^{1/2}\theta)
[dz+\epsilon \chi_{(k)}]
-k J\left[k^{-1/2}\sin(k^{1/2}\theta)\right]^{2}d\varphi\ , 
\\
&
\\
A^{3}=&
-\frac{\Psi F}{\varrho^{\epsilon}}k^{-1/2}\sin{(k^{1/2}\theta)}\sin{\varphi} 
[dz+\epsilon \chi_{(k)}]
\\ 
&
\\
&
-J
\left[\cos{\varphi}\: d\theta
- k^{-1/2}\sin{(k^{1/2}\theta)}\cos(k^{1/2}\theta)\sin{\varphi} \: 
d\varphi
\right]\ .
 \\
\end{aligned}
\end{equation}

This is our main result, but we can elaborate it a bit more.

For $\epsilon=1$ and $k\neq0$, one can also write the instanton fields in a
more compact form by using a generalization of the Maurer-Cartan forms\footnote{Here we are using a shifted coordinate $\tilde z=z-k^{-1}\varphi$.}

\begin{equation}
\begin{aligned}
v^{1}&=-\sin{\varphi}\: d\theta +k^{1/2}\sin(k^{1/2}\theta)\cos{\varphi} \:d\tilde z \ , \\
v^{2}&=d\varphi+k\cos(k^{1/2}\theta)\:d\tilde z \ , \\
v^{3}&=\cos{\varphi} \:d\theta +k^{1/2}\sin(k^{1/2}\theta)\sin{\varphi} \:d\tilde z \ , 
\end{aligned}
\end{equation}

\noindent
which satisfy $dv^{I}= -\frac{1}{2}f_{JK}{}^{I}v^{J}\wedge v^{K}$
 and in terms of which the instanton field reads

\begin{equation}
A^{I}
=
-\frac{k\Psi F}{\varrho}v^{I}
+\left(J-k\frac{\Psi F}{\varrho}\right) u^{I} \ ,
\end{equation} 

\noindent
where the $u^{I}$'s are given by 

\begin{equation}
\begin{aligned}
u^{1}
&=
\sin{\varphi} \:d\theta 
+k^{-1/2}\sin(k^{1/2}\theta)\cos(k^{1/2}\theta)\cos{\varphi} \:d\varphi \ , 
\\
& \\
u^{2}
&=
-k(k^{-1/2}\sin(k^{1/2}\theta))^{2}\:d\varphi \ , 
\\
& \\
u^{3}
&=-
\cos{\varphi} \:d\theta 
+k^{-1/2}\sin(k^{1/2}\theta)\cos(k^{1/2}\theta)\sin{\varphi} \:d\varphi \ .
\end{aligned}
\end{equation}
and satisfy $du^{I}=f_{JK}{}^I\:u^J\wedge u^K$.

Before we give an expression for the associated field strengths, we find
convenient to introduce a basis of three self-dual 2-forms

\begin{equation}
 \mathcal{B}^{i} = e^{\sharp}\wedge e^{i}+\tfrac12 \varepsilon_{jk}{}^i e^j\wedge e^k\,,
\end{equation}


\noindent
where $e^{a}$ is the Vierbein basis of the four-dimensional metric 

\begin{equation}
\begin{array}{rclrcl}
e^{\sharp} 
& = &
{\displaystyle\frac{\Psi^{1/2}}{\varrho^{\epsilon/2}}}[dz+\epsilon\chi_{(k)}]\ , 
\hspace{1cm}
&
e^{1}
& = & 
\varrho^{\epsilon/2}d\theta \ , 
\\
& & & & & \\
e^{2}
& = &
{\displaystyle\frac{\varrho^{\epsilon/2}}{\Psi^{1/2}}}d\varrho \ , 
&
e^{3}
& = & 
\varrho^{\epsilon/2}k^{-1/2}\sin(k^{1/2}\theta)\: d\varphi \, .  
\end{array}
\end{equation}

Then, in terms of these three self-dual three forms the field strengths are

\begin{equation}
\begin{aligned}
F^{1}
= &
\frac{\Psi^{1/2}J'}{\varrho^{\epsilon}} \cos(k^{1/2}\theta)\cos{\varphi}
\mathcal{B}^{1}
+\left(\frac{\Psi F}{\varrho^{\epsilon}}\right)'k^{-1/2}\sin(k^{1/2}\theta)\cos{\varphi}\mathcal{B}^{2}
\\
&
\\
&
-\frac{\Psi^{1/2}J'}{\varrho^{\epsilon}} \sin{\varphi} \mathcal{B}^{3}\ ,
\\ 
& 
\\
F^{2}
=&
-k\frac{\Psi^{1/2}J'}{\varrho^{\epsilon}}
k^{-1/2}\sin(k^{1/2}\theta)\mathcal{B}^{1}
+\left(\frac{\Psi F}{\varrho^{\epsilon}}\right)' 
\cos(k^{1/2}\theta)\mathcal{B}^{2} \ , 
\\
&
\\
F^{3}
=&
\frac{\Psi^{1/2}J'}{\varrho^{\epsilon}}
\cos(k^{1/2}\theta)\sin{\varphi}\mathcal{B}^{1}
+\left(\frac{\Psi  F}{\varrho^{\epsilon}}\right)'
k^{-1/2}\sin(k^{1/2}\theta)\sin{\varphi}\mathcal{B}^{2}
\\
&
\\
&
+\frac{\Psi^{1/2}J'}{\varrho^{\epsilon}}\cos{\varphi} \mathcal{B}^{3} \ .
\end{aligned}
\end{equation}

For later use, it is interesting to have an explicit expression for Tr$
F\wedge F\sim \eta_{IJ}F^{I}\wedge F^{J}$, even if in most cases there is no
well-defined notion of instanton number (density). One has

\begin{equation}
\begin{aligned}
\eta_{IJ}F^{I}\wedge F^{J}
& 
=
d^{4}x\, \sqrt{|g|} 
\left\{
2\left[\left(\frac{\Psi F}{\varrho^{\epsilon}}\right)'\right]^{2}
+4k\frac{\Psi (J')^{2}}{\varrho^{2\epsilon}}
\right\}
\\
& 
\\
&
=
d^{4}x\, \sqrt{|g|}\,  
\frac{2}{g^{2}\varrho^{2\epsilon}}
\left[\left(K'-\epsilon\frac{K}{\varrho}\right)^{2}
+2\frac{K^{2}}{\Psi}(K'+k)\right] 
\\
& 
\\
&
=
d^{4}x\, \sqrt{|g|}\, 
\frac{2}{g^{2}\varrho^{2\epsilon}}
\left[\left(G-1-\epsilon k\frac{K}{\varrho}\right)^{2}
+2k\frac{K^{2}G}{\Psi}\right]
\\
&
\\
&
=
\frac{2}{g^{2}}
d\left[
k\frac{(G-1)K}{\varrho^{\epsilon}}-\frac{\epsilon}{2}\frac{K^{2}}{\varrho^{\epsilon+1}}
\right]
\wedge k^{-1/2}\sin{(k^{1/2}\theta)}dz \wedge d\theta\wedge d\varphi\, .
\end{aligned}
\end{equation}

\noindent
where we have defined the functions
\begin{equation}
\label{eq:changeofvariables}
K\equiv g\Psi F\, , \hspace{1.5cm} G \equiv (1+gJ )^{2}\, ,
\end{equation}
for reasons that will become clear in the next section, and
where, in the second line, it has been assumed that $k\neq 0$.
 
For $k=1$ 

\begin{equation}
\int F^{I}\wedge F^{I}
=
\frac{8\pi T}{g^{2}}
\left[
\frac{(G-1)K}{\varrho^{\epsilon}}
-\frac{\epsilon}{2}\frac{K^{2}}{\varrho^{\epsilon+1}}\right]_{\varrho_{0}}^{\varrho_F}\,
,
\end{equation}
 
\noindent
where $T$ is the period of $z$ and $\varrho_{0},\varrho_{F}$ the limits of
integration of $\varrho$, which depend on the chosen K\"ahler space.

\subsection{Solutions}
\label{sec-solutions}

Let us now go back to the solutions of the system Eqs.~(\ref{eq:system1}). We
consider the $k=0$ and $k\neq 0$ separately.

\subsubsection{The $k=0$ case}

In this case, the first equation of (\ref{eq:system1}) can be integrated
directly for arbitrary $\Psi(\varrho)$, giving

\begin{equation}
F(\varrho) = \frac{K_{0}}{g\Psi(\varrho)}\, ,
\end{equation}

\noindent
where $K_{0}$ is an integration constant. Plugging this result into the second
equation we get

\begin{equation}
J(\varrho) =  C\: e^{\mathcal{I}(\varrho)}-\frac{1}{g}\, ,  
\end{equation}

\noindent
where $C$ is another integration constant and 

\begin{equation}
\mathcal{I}(\varrho) 
\equiv
K_{0}\int^{\varrho} \frac{du}{\Psi(u)}\, .
\end{equation}

For instance, the metric of $\overline{\mathbb{CP}}^{2}$ can be written in the
$k=0$ form with $\Psi =4\varrho^{3}/\ell^{2}$ and $H=\varrho/\Psi$,
(\textit{i.e.}~$\epsilon=1$) \cite{Chimento:2016mmd} and, therefore, we have

\begin{equation}
F(\varrho)=\frac{\lambda}{g\varrho^{3}}\, , 
\hspace{2cm}
J=C e^{-\frac{\lambda}{2\varrho^{2}}} -\frac{1}{g}\, .
\end{equation}

\subsubsection{The $k\neq 0$ case}

The system Eqs.~(\ref{eq:system1}) can be simplified with the change of
variables (\ref{eq:changeofvariables}), after which it takes the form

\begin{equation}
\label{eq:systemBog}
\begin{aligned}
K'
& = 
k (G-1)\, ,
\\
\Psi G'
& = 
2KG\, .
\end{aligned}
\end{equation}

The first equation in (\ref{eq:systemBog}) can be used to eliminate $G$ in the
second one, which leads to a second order equation that only involves the
variable $K$:

\begin{equation}
\label{eq:2order}
\Psi K''-2KK'-2kK=0\, .
\end{equation}

Given the function $\Psi(\varrho)$ corresponding to a K\"ahler metric in the
class we are considering, this equation determines $K$. Observe that we can
turn around the problem and choose some arbitrary $K(\varrho)$ and then find
the K\"ahler manifold in which it defines a selfdual instanton by computing
directly

\begin{equation}
\Psi = \frac{2K(K'+k)}{K''}\, .  
\end{equation}

There is a simple solution to Eq.~(\ref{eq:2order})\footnote{This solution
  is also available in the $k=0$ case and corresponds to the choice of the
  integration constant $C=0$.}  which is valid for any $\Psi$, with $K''=0$:
$K=K_{0}-k\varrho$, $G=0$. The functions $F$ and $J$ that appear in the
hedgehog ansatz Eq.~(\ref{eq:hedgehogansatz}) are given by\footnote{In this
  case $k=1/k$ and, in the form in which we are giving this general solution,
  it is automatically valid for the $k=0$ case.}

\begin{equation}
F(\varrho)=\frac{K_{0}-k\varrho}{g\Psi(\varrho)}\, , 
\hspace{1cm}
J=-\frac{1}{g}\, .
\end{equation}

This solution corresponds to a fixed point of the system
Eqs.~(\ref{eq:systemBog}).

In order to find more solutions we need to know $\Psi(\varrho)$. In many
interesting cases $\Psi(\varrho)$ is a polynomial of order $N$, and, in
particular, with $N=3$.  If $\Psi(\varrho)$ is a polynomial of order $N$
we can assume that $K$ is also a polynomial whose order must be
$N-1$ for Eq.~(\ref{eq:2order}) to have solutions, in general. The
differential equation becomes a set of algebraic equations relating the
coefficients of the polynomial $K$ to those of the polynomial $\Psi$. For
$N=3$, if

\begin{equation}
\begin{aligned}
\Psi(\varrho) 
& =
\Psi_{0} +\Psi_{1}\varrho+ \Psi_{2}\varrho^{2}+\Psi_{3}\varrho^{3}\, ,
\\
K(\varrho) 
& =
K_{0}+K_{1}\varrho+K_{2}\varrho^{2}\, ,
\\
\end{aligned}
\end{equation}

\noindent
one readily finds the following relations ($\Psi_{3}\neq 0$, by assumption)

\begin{equation}
\label{eq:coefficients}
\begin{aligned}
K_{2} 
& = 
\frac{\Psi_{3}}{2}\, , 
\\
K_{1}
& = \frac{\Psi_{2}-k}{3}\, ,
\\
K_{0} 
& = 
\frac{9\Psi_{1}\Psi_{3}-2(\Psi_{2}+2k)(\Psi_{2}-k)}{18\Psi_{3}} \, ,
\end{aligned}
\end{equation}

\noindent
and one constraint for the coefficients of $\Psi(\varrho)$ 

\begin{equation}
9\Psi_{1}\Psi_{3}(\Psi_{2}+2k)-2(\Psi_{2}+2k)^{2}(\Psi_{2}-k)=27\Psi_{0}\Psi_{3}^{2} \, ,
\end{equation}

\noindent
which has to be understood just as the condition that $\Psi$ has to satisfy in
order for Eq.~(\ref{eq:2order}) to admit a solution in which $K$ is a second
order polynomial.

Since $J$ is a real function, the second of Eqs.~(\ref{eq:changeofvariables})
$G$ must be a positive definite function. The first of
Eqs.~(\ref{eq:systemBog}) and Eqs.~(\ref{eq:coefficients}) tell us that it is
given by 

\begin{equation}
G 
= 
1+kK' 
= 
1+kK_{1} +2k K_{2}\varrho 
= 
\frac{2 +k\Psi_{2}}{3} +k\Psi_{3}\varrho\, ,  
\end{equation}

\noindent
so that $\varrho$ is restricted to the interval

\begin{equation}
\begin{aligned}
 \varrho &> -\frac{k(2 +k\Psi_{2})}{3\Psi_{3}}\,\qquad \mbox{for}\quad k\Psi_{3}>0\,,\\
 \\
 \varrho &< -\frac{k(2 +k\Psi_{2})}{3\Psi_{3}}\,\qquad \mbox{for}\quad k\Psi_{3}<0\,.
\end{aligned}
\end{equation}

The metric of $\overline{\mathbb{CP}}^{2}$ can be written in the $k=\pm 1$
form with a $\Psi$ which is the cubic polynomial

\begin{equation}
\Psi = \varrho^{2}(k+4\varrho/\ell^{2})\,   
\end{equation}

\noindent
and, since $\Psi$ must also be positive in the metric, the variable $\varrho$
is restricted to

\begin{equation}
\varrho > -k\ell^{2}/4\, ,  
\end{equation}

\noindent
which is compatible with the restriction found above only for $k=1$.

\section{Conclusions}
\label{sec-conclusions}

We have completed our program of finding simple equations that selfdual
instanton solutions in K\"ahler spaces with one holomorphic isometry have to
satisfy, generalizing Kronheimer's work in the hyperK\"ahler case. We have also
constructed some explicit solutions in some K\"ahler spaces of particular
interest from the point of view of 5-dimensional Abelian-gauged supergravity.
In passing, we have generalized the hedgehog ansatz to some non-spherical
symmetries and gauge groups different from SU$(2)$. There is little work in
the literature on non-compact gaugings and we think these results will allow
us to find interesting solutions in those cases. 

We have not analyzed the regularity of the solutions we have obtained because,
ultimately, they are going to be part of a complicated 5-dimensional gauge
field defined in a 5-dimensional spacetime whose regularity does not depend on
the regularity of each of its building blocks. There are perfectly regular
5-dimensional solutions (microstate geometries) built over singular base
spaces like ambipolar GH spaces \cite{Ramirez:2016tqc, Avila:2017pwi}. And most 
regular, charged, extremal black
holes use Coulomb-like 1-form fields which are singular at a point which, in
the end, turns out to be not a point but a regular horizon. The same mechanism
saves the regularity of the 4-dimensional non-Abelian black holes which bear
BPS magnetic monopole fields different from the 't~Hooft-Polyakov one. Thus,
one should not be worried about the possible singularities of the instantons
until the full 5-dimensional supergravity solution is constructed. For the
same reason (and also because the K\"ahler spaces that we are considering are
not compact) we have not computed the instanton number.

In a forthcoming publication \cite{ChOR} we will put to use the results
obtained in this work, which we hope will also be useful in other contexts.

\section*{Acknowledgments}

The authors would like to thank Pedro F. Ram\'irez for his initial collaboration 
in this work and for interesting discussions.
This work has been supported in part by the MINECO/FEDER, UE grant
FPA2015-66793-P and by the Spanish Research Agency (Agencia Estatal de
Investigaci\'on) through the grant IFT Centro de Excelencia Severo Ochoa
SEV-2016-0597.  TO wishes to thank M.M.~Fern\'andez for her permanent support.

\appendix


\end{document}